\documentstyle[aps,prl,psfig]{revtex}
\begin{document}
\draft
\title{Critical Thickness in Dewetting Films}
\author{B.\ Du, Z.\ Yang, and O.\ K.\ C.\ Tsui \footnote{ To whom correspondence should be addressed.
Electronic address: phtsui@ust.hk}}
\address{Institute of Nano Science and Technology and Department of Physics, Hong Kong University of Science and Technology, Clear Water Bay, Kowloon, Hong Kong.}
\maketitle
\date{\today}
\begin{abstract}
We study
dewetting of thin polymer films with built-in topographical fluctuations produced 
by rubbing the film surface with a rayon cloth. By varying the density of 
imposed surface defects, we unambiguously
distinguish spinodal dewetting, which dominates in liquid films thinner than 
a characteristic thickness $= 13.3$~nm, from heterogeneous nucleation in
the thicker films. Invariance of this characteristic
thickness upon more than a decade change in 
the defect density makes kinetic effect an unseemly origin.
A crossover of the spinodal line provides a consistent picture.
This interpretation, however, contends the current understanding of molecular 
interactions in apolar liquid films.
\end{abstract}

\twocolumn

A physical state is unstable if the curvature of its free energy, $G$ with respect to the order parameter, $h$ is negative. According to the mean-field theory
of Cahn \cite{Cahn}, 
such states may proceed to equilibrium via a spinodal mechanism that results in  a spatial modulation of the order parameter. It can be shown that the mode with wavelength, 
$\lambda_m = (-4\pi^2\gamma/d^2G/dh^2)^\frac{1}{2}$ will grow the fastest, 
where $\gamma$ is the rigidity of the system order parameter against 
deviations from homogeneity. Alternatively, the system may evolve to equilibrium
by nucleation of heterogeneity at defect sites from within. 
For a long time, the distinction between these two 
mechanisms in the dewetting of liquid films from a non-wetting
substrate has been the subject of many contentious debates \cite{Reiter_Langmuir99,Xie98,Reiter_Europhys99,Meredith2000,ReiterPRL,Sharma96,Sharma_PRL98,Jacobs98,Herminghaus_PRL96,Thiele98}. 

In supported apolar liquid films, 
the non-retarded Lifsitz-van der Waals interactions render the unit
area free energy the form $-A/12\pi h^2$ \cite{Israelachvili}, where $A$ is
the Hamaker constant and $h$ - the system 
order parameter - is the liquid film thickness. Therefore, 
thin liquid films wherein $A > 0$ will rupture spontaneously by
spinodal growth of surface capillary waves, for which the fastest growing mode has
the wavelength, $\lambda_m =  
(16\pi^3\gamma/ A)^\frac{1}{2}h^2$ \cite{Sharma96,Vrij,Brochard} (Here, $\gamma$ is the surface tension of the liquid.) 
Previous experiments find that
dewetting of liquid films proceed according to a generic sequence \cite{Reiter_Langmuir99,Xie98,Reiter_Europhys99,Meredith2000,ReiterPRL,Sharma96,Sharma_PRL98,Jacobs98} beginning with a thickness modulation,
often noted as holes, followed by the coarsening of the holes and
formation of a polygon network, with the eventual breakup of
the polygon network into droplets due to the Rayleigh instability \cite{Rayleigh}.
While the latter processes are governed by hydrodynamics of dewetting and 
well understood \cite{Brochard,Rayleigh,Redon}, 
opinions are divided on the mechanism (i.e.\ spinodal
or heterogeneous nucleation) underlying the crucial initial step.
An $h^{-2}$ dependence 
frequently found in the dewetting wavevector \cite{Reiter_Europhys99,Meredith2000,ReiterPRL,Sharma96,Herminghaus_PRL96}
has been
taken as evidence for spinodal dewetting. 
However, Jacobs et al.\ \cite{Jacobs98} 
found the holes to have no positional correlation, thereby unlikely 
a result of spinodal surface undulations. In addition,
the $h^{-2}$ dependence 
apparent in the experimental characteristic wavevector has also been argued to
resemble an exponential variation $\sim \exp(-h/L)$ ($L=$ constant), 
fitting to be considered an evidence of 
heterogeneous nucleation from intrinsic defects. 
Furthermore, it was found in dewetting experiments of metal thin films 
\cite{Herminghaus_PRL96} and
evaporating volatile liquid films \cite{Thiele98} that morphologies 
due to both kinds of 
mechanisms may sometimes coexist. This adds further complications 
to the attempt to make a clean distinction between the two destabilizing
mechanisms using conventional approaches
that rely on direct analysis of the dewetting patterns.

In this experiment, we used polystyrene (PS) spin-coated on silicon covered with
an oxide layer as the model system, which is the most studied for this problem.
We artificially introduce small height fluctuations, $\delta h$ 
$(\leq 1$~nm) to the thin film samples 
by mechanically rubbing the sample surface with a rayon cloth before
dewetting initiates. For liquid films that dewet by nucleation,
the characteristic length of the
dewetting pattern depend on the density of 
surface fluctuations introduced. 
But for those films that dewet by a spinodal process, the characteristic length
should remain unaffected \cite{RDS}. Therefore, simply by 
comparing the dewetting 
morphologies with and without the rubbing-induced surface defects,
one can unambiguously 
identify which of the two dewetting
mechanisms dominates. 
Our results show that the final pattern of all 
dewet films is always dictated by {\em just one}
mechanism. 
For the present system (PS/SiO$_2$/Si), it is found that 
spinodal dewetting dominates in the thinner films ($h < 13.3$~nm) whereas 
heterogeneous nucleation dominates in the thicker ones. Abruptness of the
transition at $h = 13.3$~nm between the two mechanisms suggests
that it is of a thermodynamic rather than kinetic nature. We attribute
the transition to a crossover of the spinodal line, which however may 
present challenge to our understanding of molecular interactions 
for the present liquid film system. We will elaborate further on this point 
below. By examining the dewetting morphologies in different 
regions with respect to the critical thickness,
we are able to make direct comparison between experiment 
and theoretical predictions \cite{Sharma_PRL98}
on dewetting pattern formation. 
Excellent agreement between theory and experiment is obtained. 
Our results thus resolve the controversy by showing that the two
mechanisms are operative in different thickness regions separated
by a transition.

The PS sample was purchased
from Scientific Polymer Products (Ontario, NY). It has a
molecular weight of 13.7K~Da and polydispersity 1.1.
The glass transition temperature was measured to be $\sim 99~^\circ$C
by differential scanning calorimetry. 
The substrate overlayer was prepared by wet oxidation: 4" diameter 
Si(100) wafers 
cleaned as described previously \cite{cleaning} were annealed at 
1000~$^\circ$C in a reaction chamber filled with water vapor (produced 
by a simultaneous injection of 5.6~L/min.\ H$_2$ and~4 L/min.\ O$_2$ into
the chamber) 
for 9~mins.~40~sec. The thermal oxide was confirmed  
to be 106~nm thick by spectroscopic ellipsometry with better than 1 \% uniformity.
The oxidized wafers were then cut into square pieces of $1 \times 1$~cm$^2$, 
and cleaned \cite{cleaning} before coated with the polymer. Both the
substrate cleaning and polymer coating were carried out in a clean
room. Before use, the polymer films were annealed at $100~ ^\circ$C
under vacuum ($10^{-2}$~torr) for 5~h to remove the residue solvent.
No sign of dewetting was detectable in the samples 
after annealing. To produce the artificial topographical fluctuations, 
a piece of rayon cloth loaded with a
normal pressure of 10~g/cm$^2$ was rubbed against the film surface 
at a constant speed = 1~cm/s by an 
electric motor. The
density of surface defects was varied by changing the number of
rubs applied to the films. Sample topography was characterized
by a Seiko Instruments (Chiba, Japan) SPA-300HV atomic force microscope (AFM).  
Fig.~\ref{fig:freshrub} shows the Fourier spectra for the surface topography
of five freshly rubbed PS films that had been rubbed with different number 
of times ($N = 3$ to 15).
These spectra were obtained by radial averaging the
two-dimensional (2D) fast Fourier transformation (FFT) of 
AFM topographical 
images ($5\times 5~\mu$m$^2$) obtained from the samples. As seen,
all spectra look alike. Except for a gentle, but obivious rise
towards $q = 0$ and a small peak at $q = 120$~$\mu$m$^{-1}$, the
spectra are overall speaking rather flat. From the 
FFT images (for example, left inset of Fig.~\ref{fig:freshrub}), one can
see that the spectra are dominated
by anisotropic topographies parallel to the rubbing direction.
We found that the film roughness could be fitted to $(0.26^2 + 0.16^2N)^\frac{1}{2}$ (nm)
(solid line in the right inset of Fig.~\ref{fig:freshrub}). All these
characterizations about the topography of the rubbed films
suggest that addition of height fluctations in the films by individual
rubs are independent events. It thus follows that the
density of rubbing-induced surface defects should increase
linearly with $N$. 

Dewetting experiments were carried out in air. The films were annealed at
180~$^\circ$C for 10 to 60~mins. We deduce the characteristic wavevector,
$q^\ast$ of the final dewetting patterns from the peak of
their radial averaged Fourier spectra. 
When the initial holes do not coalesce to form a network of polygons
before break up into droplets, which has been found in thinner films where
$h < 13.3$~nm, $(q^{\ast})^2$ will be $\sim N_d$, 
the areal density of the final 
liquid droplets \cite{Reiter_Europhys99}. For the thicker ones,
$(q^{\ast})^2$ is $\sim N_p$, the areal density 
of the polygons \cite{Meredith2000,ReiterPRL,Sharma96}. 
Fig.~\ref{fig:qvsh} depicts $q^{\ast}$ vs.\ $h$ for 
samples rubbed by different number of times ($N = 0$ to 10) in a 3D plot. As seen,
the data are divided into two groups at $h = 13.3$~nm: Below 13.3~nm,
$q^\ast$ is independent of $N$; but above, it increases with increasing $N$.
This result unambiguously evidences that samples with $h < 13.3$~nm rupture by a
spinodal mechanism whereas those with $h > 13.3$~nm rupture by nucleation 
of holes. We note that $q^\ast(h)$ of the unrubbed films fits quite well to
$\sim h^{(-2.1 \pm 0.05)}$ for $h < 13.3$~nm, and $\sim h^{(-2.5 \pm 0.1)}$  
for $h > 13.3$~nm (solid lines in Fig.~\ref{fig:qvsh}). The fact that the
latter is not notably different from the $h^{-2}$ dependence could 
have been easily mistaken
as the signature of spinodal dewetting. We have
model-fit the data to an exponential function 
(data not shown) and obtained
similar good fits as Jacobs et al.\ found
\cite{Jacobs98}. This finding confirms the earlier suggestion by 
these authors
that the $q^\ast \sim h^2$ scaling characteristic is insufficient
for the distinction between spinodal and nucleation dewetting. 

We now examine evolution of the dewetting morphology of unrubbed PS 
films in different thickness regimes about the threshold at $h=13.3$~nm. 
Fig.~\ref{fig:6.8nm} shows AFM topographical images obtained 
from a 6.8~nm thick PS film upon quenching at different annealing
times from 3~sec.\ to 23~min.\ at 145~$^\circ$C. As seen, uniformly sized 
holes first appeared after 10 sec., which enlarged and 
developed
into a bicontinuous structure after 7~mins. In Fig.~\ref{fig:13nm}, we contrast evolution 
of the rupturing morphology at 145~$^\circ$C of PS films near the threshold: one just below at $h=12.8$~nm and one 
just above at $h=13.8$~nm. For the $h=12.8$~nm thick PS film (Fig.~\ref{fig:13nm}(a)), 
non-uniformly sized holes were first found scattering randomly across the sample 
after 4~mins. They grew in size, and more holes emerged
with time. Until $\sim 28$~mins.\, the holes filled up the 
entire area of the film whereupon the thin ribbons connecting the holes 
broke up into droplets. 
The $h=13.8$~nm thick PS film, though differ in thickness from the $h=12.8$~nm 
one by only 1~nm, undertook a drastically different pathway (Fig.~\ref{fig:13nm}(b)). 
As seen, randomly distributed holes also
appeared in the first few minutes, but their size and number never grew
large enough to cover up the entire film at similar times it had taken the $h=12.8$~nm
thick film. Instead, only bigger holes with a much lower areal coverage could be seen
growing and emerging, which after a much longer time of $\sim 100$~mins., 
finally touch each other and coalesced into polygons
before the thin ribbons connecting the polygons broke up into droplets.

With illustrations of Figs.~\ref{fig:6.8nm} and \ref{fig:13nm}, we may summarize
on what kinds of conformation can be formed in the early stage of spinodal 
dewetting of 
supported liquid films. Basically, both a bicontinuous structure 
(Fig.~\ref{fig:6.8nm}) and distribution of circular holes (Fig.~\ref{fig:13nm}(a)) can  
form in the initial stage of a spinodal process.  Our results show
that the bicontinuous structure is formed only in samples in the deep spinodal
region whereas circular holes are formed in samples closer to the spinodal line,
consistent 
with the 3D nonlinear calculations of Sharma et al.\ \cite{Sharma_PRL98}. 
It is, however, noteworthy that the linear theory \cite{Cahn} will have
predicted only the formation of a bicontinuous structure. Our result, therefore, 
clearly demonstrates
inadequacy of the linear theory in accounting for the spinodal process.
Another point to note is that the circular holes formed generally do 
not have positional correlations (see Fig.~\ref{fig:13nm}(a)) although the 
charcteristic $q^\ast$ is still 
the same as the one predicted by the 
linear theory \cite{Cahn,Vrij,Brochard}, which is again in consistency with 
Ref.~\cite{Sharma_PRL98}. 

It remains to understand what causes the transition between different 
dewetting mechanisms 
near $h=13.3$~nm. Similar transition
has also been observed in previous experiments 
\cite{Xie98,Meredith2000}. Meredith et al.\ attributed the transition
to a crossover between spinodal and nucleation process due to 
kinetic dominance of the latter over the former 
at large $h$ \cite{kinetic_effect}.  We note that the 
critical thickness remains unchanged even when the density of rubbing-induced defects 
is varied by more than an order of magnitude. Kinetic effects will require that 
the critical thickness should change by more than a factor 
$\sim 10^{\frac{1}{5}} (= 1.58)$ \cite{kinetic_effect}, which is obviously
not evident in the data of Fig.~\ref{fig:qvsh}. Therefore, it is most
probable that the transition represents a crossover of the spinodal line. 
This interpretation, however, necessitates that the free energy has an
inflection point at $h \approx 13.3$~nm. Using the more precise form of the free energy
proposed in Ref.~\cite{Sharma96} to take into account that the present system, is
a PS/SiO$_2$/Si tri-layer rather than a simple bilayer one for which the $G \sim
A/h^2$ relation is derived, one will obtain a prediction for the position of
the spinodal line to be at $h \geq \delta = 106$~nm, the thickness of
SiO$_2$ in this experiment. At present,
the reason for the discrepancy between the predicted and the observed
transition at $h=13.3$~nm is not entirely clear. It is though apparent that
retardation effects may no longer be negligible for the
Lifsitz-van der Waals interactions across
the SiO$_2$ layer since $\delta \approx 100$~nm is quite large in this study 
\cite{Israelachvili}. However, retardation effects will more likely 
worsen the discrepancy between the predicted and the experimental
critical thickness. Studies are currently being
carried out to understand how the SiO$_2$ inter-layer
thickness affects the position of the critical thickness. It is
hoped that with the insight thus gained
about the form of van der Waals interactions in a tri-layer system, 
stability of thin liquid films can be premeditated.

We thank R.\ Seemann, K.\ Jacobs and S.\ Herminghaus for sending us their 
manuscript prior to publication. Critical reading of this manuscript by 
P.\ Sheng is gratefully acknowledged. This work is 
supported by the Institute of Nano Science and Technology, HKUST.


%
%
\begin{figure}
\caption{Fourier spectra of surface topography of PS films rubbed with different
number of times. Left inset shows FFT image obtained from an AFM topographical
image of a PS film that was rubbed 10 times. Right inset shows the roughness
of rubbed PS films vs.\ number of rubs. Filled circles are data. Solid line
is the model fit (see text).}
\label{fig:freshrub}
\end{figure}

\begin{figure}
\caption{3D plot of the characteristic wavevector, $q$\* vs.\ $h$ for 
PS films rubbed
by different number of times from 0 to 10. Solid lines are fits to a power
law. Thin dotted lines are contour lines at $h = 7, 13.3$ and 30~nm.}
\label{fig:qvsh}
\end{figure}

\begin{figure}
\caption{AFM topographical images of a 6.8~nm thick PS film taken
after quenching at different annealing times from
10~sec.\ to 23~mins.\ at 145$^\circ$C.}
\label{fig:6.8nm}
\end{figure}

\begin{figure}
\caption{Comparison between dewetting morphology of (a) 12.8~nm and (b) 13.8~nm 
thick PS films at 145$^\circ$C. All pictures are AFM topographical 
images except the last two
in (b), which are optical micrographs.}
\label{fig:13nm}
\end{figure}

%
%
%

\end{document}